\documentclass[%  
reprint,superscriptaddress,
amsmath,amssymb,
aps,
prb,
showpacs,
]{revtex4-1}
\usepackage{color}
\usepackage[usenames,dvipsnames,svgnames,table]{xcolor}
\usepackage{amssymb}
\usepackage{mathrsfs}
\usepackage{graphicx}% Include figure files
\usepackage{dcolumn}% Align table columns on decimal point
\usepackage{bm}% bold math
\usepackage{hyperref}% add hypertext capabilities
\hypersetup{colorlinks=true,linkcolor=NavyBlue,citecolor=BrickRed,urlcolor=Blue}

\begin{document}

\providecommand{\abs}[1]{\lvert #1 \rvert}
\providecommand{\Ham}{\mathcal{H}}
\providecommand{\bfmath}[1]{\mathbold #1 }
\newcommand\blue[1]{{\color{blue}#1}}
\newcommand\red[1]{{\color{red}#1}}
\newcommand\org[1]{{\color{orange}#1}}

\title{Topological surface states in strained Dirac semimetal thin films}

\author{Pablo Villar Arribi}
\affiliation{Materials Science Division, Argonne National Laboratory, Lemont, Illinois 60439, USA}
%\email{pvillararribi@anl.gov}
%\author{Zhao Huang}
%\affiliation{Theoretical Division, Los Alamos National Laboratory, Los Alamos, New Mexico 87545, USA}
%\affiliation{Center for Integrated Nanotechnologies, Los Alamos National Laboratory, Los Alamos, New Mexico 87545, USA}
\author{Jian-Xin Zhu}
\affiliation{Theoretical Division, Los Alamos National Laboratory, Los Alamos, New Mexico 87545, USA}
\affiliation{Center for Integrated Nanotechnologies, Los Alamos National Laboratory, Los Alamos, New Mexico 87545, USA}
\author{Timo Schumann}
\affiliation{Materials Department, University of California, Santa Barbara, CA 93106}
\author{Susanne Stemmer}
\affiliation{Materials Department, University of California, Santa Barbara, CA 93106}
\author{Anton A. Burkov}
\affiliation{Department of Physics and Astronomy, University of Waterloo, Waterloo, Ontario, Canada N2L 3G1}
\author{Olle Heinonen}
\email{heinonen@anl.gov}
\affiliation{Materials Science Division, Argonne National Laboratory, Lemont, Illinois 60439, USA} 

\date{\today}
\begin{abstract}
We computationally study the Fermi arc states in a Dirac semimetal, both in a semi-infinite slab and in the thin-film limit. We use Cd$_3$A$_2$ as a model system, and include perturbations that break the $C_4$ symmetry and inversion symmetry. The  surface states are protected by the mirror symmetries present in the bulk states and thus survive these perturbations. %no matter which of these two perturbations occurs. 
The Fermi arc states persist down to very thin films, thinner than presently measured experimentally, but are affected by breaking the symmetry of the Hamiltonian. Our findings are compatible with experimental observations of transport in Cd$_3$As$_2$, and also suggest that symmetry-breaking terms that preserve the Fermi arc states nevertheless can have a profound effect in the thin film limit.
\end{abstract}

\maketitle
%\section{Introduction}\label{sec:intro}

%\textcolor{blue}{Start with brief introduction of topologically non-trivial materials: TIs, Dirac, and Weyls and something about the relation between them. Next introduce the surface states and Fermi arc states and why they are interesting - this should include also QHE in addition to other recent effects utilizing topological surface states.} 

The discovery of topologically non-trivial phases of matter has revolutionized the world of solid-state physics and materials science in the past decade. These systems exhibit new exotic electronic properties that are of great fundamental interest and also make them versatile for potential technological applications. Among these phases, the most important ones to date are topological insulators (TIs), Dirac semimetals (DSMs), and Weyl semimetals (WSMs)\cite{Armitage_RevModPhys.90.015001,Wang_PRB2012, Young_3D_DSM_PRL2012, Wan_PRB2011, Xu_Chern_SM_PRL_2011, Lv_PRX2015}. A peculiar property of these systems is that they exhibit unusual surface states that are protected by symmetry or topology against perturbations\cite{Hasan_Kane_RMP, Bansil_RMP, Kane_Mele_Z2_PRL95, bernevig_science_2006, FU_TI3D_PRL_2007, Moore_Balents_PRB_2007}. Also common to TIs, DSMs, and WSMs is that spin-orbit coupling is relatively strong and leads to band inversion in parts of the Brillouin zone. %The peculiarity of these systems is that they display different properties in the bulk than in their surface, where the emerging electronic states are, in general, much more robust against perturbations than their bulk counterparts. 

%\textcolor{magenta}{Introduce the surface states and Fermi arc states and why they are interesting - this should include also QHE in addition to other recent effects utilizing topological surface states.} 

%TIs are materials which are insulating in the bulk %in which the quasi-particle excitations are gapped in the bulk (insulators) 
%but have gapless surface states, usually called topological surface states (TSS), that are %topologically 
%protected by time-reversal symmetry.\cite{Hasan_Kane_RMP, Bansil_RMP, Kane_Mele_Z2_PRL95, bernevig_science_2006, FU_TI3D_PRL_2007, Moore_Balents_PRB_2007} The TSS of three-dimensional TIs also exhibit a Dirac dispersion, analogous to the two-dimensional semimetal graphene. %, and are chiral with specific relations between spin and momentum.

DSMs have special points in the Brillouin zone where the valence and conduction bands touch. These points, known as Dirac points, are protected both by time-reversal symmetry (TRS) and inversion symmetry. Because TRS is respected, there is a double Kramers degeneracy at the Dirac points, and near them, the electron dispersions are approximately described by the massless Dirac equation. The surface states in a DSM connect pairs of Dirac points in so-called Fermi arcs which, similar to the TI surface states, have specific spin-momentum locking.

%A WSM also has special points the Brillouin zone where the conduction and valence bands touch. In contrast with the DSM, a WSM has broken either TRS or inversion symmetry; indeed, a WSM can be obtained from a DSM by breaking one of those two symmetries which causes the degeneracy at each Dirac point to be lifted and the band touching points, the Weyl nodes, to move apart in the Brillouin zone. Each Weyl node is a source of Berry flux and can give rise to appreciable magnetotransport responses, such as the anomalous Hall effect (AHE)\cite{ye1999berry,taguchi2001spin,jungwirth2002anomalous,nagaosa2010anomalous}. %In contrast to the surface states in a TI or the Fermi arc states in a DSM that are protected only by symmetry, the Fermi arc states in a WSM enjoy a stronger topological protection\cite{Armitage_RevModPhys.90.015001}.

The surface states in topological materials are of great interest since they have explicit coupling between momentum and spin (spin-momentum locking) which has as a consequence that transport in the surface states is dissipationless. Furthermore, the spin-momentum locking can be used to impart spin Hall torque or used in other spintronics applications~\cite{miron2011perpendicular, liu2012spin, khang2018conductive, mahendra2018room,Zhang_Heinonen_PhysRevLett.123.187201,li2019magnetization}, and magnetic WSMs with broken TRS can give rise to a high-temperature quantum anomalous Hall effect\cite{liu2008quantum,chang2013experimental}, which is of intense fundamental and practical interest. Because of the unique properties and great versatility of the Fermi arc states, it is important to understand how they respond to external perturbations, such as strain. Furthermore, quantum transport such as the quantum Hall effect or the quantum anomalous Hall effect is manifested when the bulk states do not contribute to the transport\cite{liu2008quantum,chang2013experimental,Schumann_PhysRevLett.120.016801}. This can in principle be achieved by gapping out the bulk states by going to the extreme thin-film limit where quantum confinement leads to a discrete spectrum of gapped bulk states.  However, Fermi arc states decay exponentially into the bulk with a decay length that diverges at the Dirac or Weyl nodes. As a DSM or WSM thin film is made thinner, the Fermi arc states on the two surfaces will couple. This brings up a very important question: what happens with with the Fermi arc states in the thin film limit?

In our work, we use Cd$_3$As$_2$ as a model. This compound features a three-dimensional linear excitation spectrum near the Fermi level~\cite{Borisenko_PhysRevLett.113.027603,yi2014_sci_rep,liu_nat_mater_2014}, high mobility carriers~\cite{rosenberg_aip_1959, neupane_nat_comm_2014}, and it turns out to be prototypical DSM~\cite{Crassee_PhysRevMaterials.2.120302}: It only has two Dirac nodes near the $\Gamma$ point in the first Brillouin zone, very close in energy to the Fermi surface, and the line connecting them is parallel to the crystallographic $c$-axis. The structure of Cd$_3$As$_2$ is rather complicated with a large unit cell. Thin films of this material can be grown along the [112] direction (which yields a projection of the two Dirac nodes on the (112) surface) over different substrates with a similar lattice parameters~\cite{Schumann_APL4_MBE_Cd3As2, Goyal_APL6_Cd3As2_thin_films}. This induces strain that breaks $C_4$ symmetry around the crystallographic $c$-axis that protects the Dirac bulk states while inversion symmetry is preserved \cite{pardue2020}. However recent experiments suggest that Fermi arc states are present in these thin films ~\cite{Goyal_PhysRevMaterials.3.064204, Galletti_PhysRevB.97.115132, Galletti_PhysRevMaterials.2.124202, Schumann_PhysRevLett.120.016801,zhang2017_nat_comm, uchida2017_nat_comm}, which brings up the question of how the Fermi arc states evolve from the semi-infinite case into the ultra-thin film limit in the presence of broken symmetries. 

%Here, we are using a simple description of the low energy physics in Cd$_3$As$_2$ to examine the behavior of its Fermi arc states.
%Although the electronic structure of Cd$_3$As$_2$ can be well described by usual density functional theory (DFT) calculations ~\cite{conte_sci_rep_2017, Roth_PhysRevB.97.165439}, the large number of electrons in the unit cell (80) makes a systematic and thorough study of the evolution of the electronic structure with different parameters and in the thin-film limit a rather expensive computational task. A good alternative are low-energy effective models, that can be derived from DFT or using other methods, such as the $k\cdot p$-method~\cite{Wang_PhysRevB.88.125427,jeon_nat_mater_2014,Kane_Mele_model_graphene}. These low-energy models turn out to be very useful for exploring the details of the electronic structure around the Fermi level: they can provide useful insight on the physics, they can be easily integrated into transport models, and can be used in the thin-film limit where using DFT to calculate Fermi arc states is very tedious. We find that Fermi arc states survive to very thin films (thickness $\sim$10~{\AA}) but symmetry-breaking terms in the Hamiltonian that preserve the Fermi arc states can give rise to rather striking effects in the thin-film limit, effects that in principle are observable.

%\section{Methods}\label{sec:methods}
We model the energy dispersion relation around the $\Gamma$ point with the following $4\times4$ Kane-Mele~\cite{Kane_Mele_model_graphene} Hamiltonian from Wang et al.~\cite{Wang_PhysRevB.88.125427} that has been previously used in this material:~\cite{Wang_PhysRevB.88.125427,jeon_nat_mater_2014} %but with the addition of a hyperbolic dispersion along $k_z$ (S. Jeon et al, Nature Materials) :
\begin{equation}
H_1(\mathbf{k})=\epsilon_0({\mathbf k})\mathbf{I}_4+H_2(\mathbf{k}),
    \label{eqn:H4}
\end{equation}
where $\mathbf{I}_4$ is the $4\times4$ identity matrix,  
\begin{equation}
H_2(\mathbf{k})=
    \left(
    \begin{array}{cccc}
      M({\mathbf k})   & Ak_+ & Dk_+ &  B^*(\mathbf{k})  \\
      Ak_- & -M({\mathbf k}) & B^*(\mathbf{k}) & 0 \\
      Dk_- & B(\mathbf{k}) &  M({\mathbf k}) & -Ak_- \\
      B(\mathbf{k}) & 0 & -Ak_+ & -M({\mathbf k}) 
    \end{array}
    \right),\\
    \label{eqn:H4_2}
\end{equation}
%\begin{equation}
$
    \epsilon_0({\mathbf k})= C_0+C_1k_z^2+C_2(k_x^2+k_y^2),
%    \label{eqn:c}
%\end{equation}
$
%\begin{equation}
%    M({\mathbf k})=M_0+\sqrt{M_3^2+M_1k_z^2}+M_2(k_x^2+k_y^2).
$
    M({\mathbf k})=M_0+M_1k_z^2+M_2(k_x^2+k_y^2),
    $
    and 
%    \label{eqn:M}
%\end{equation}
%\begin{equation}
$
    k_{\pm}=k_x\pm ik_y.
    $
%\end{equation}
%and
%\begin{equation}
%$
%    B({\mathbf k})=b_1 k_z$.
   % 
%\end{equation}
The Hamiltonian in Eq.~(\ref{eqn:H4}) is written in a total angular momentum representation using a basis set $|S_{\frac{1}{2}},\frac{1}{2}\rangle$, $|P_{\frac{3}{2}},\frac{3}{2}\rangle$, $|S_{\frac{1}{2}},-\frac{1}{2}\rangle$, $|P_{\frac{3}{2}},-\frac{3}{2}\rangle$ times a plane wave $\exp(-i\mathbf{k}\cdot\mathbf{r})$ valid near the $\Gamma$ point in the first Brillouin zone. This representation uses a coordinate system coincident with the tetragonal crystallographic axes $a,a$, and $c$. The parameter $D$ in Eq.~(\ref{eqn:H4_2}) breaks inversion symmetry, and we use here $B(\mathbf{k})= b_1k_z$ which breaks the $C_4$ symmetry about the tetragonal $c$ axis, as would a strain in the plane perpendicular to the $c$ axis. We note that higher-order terms in $B(\mathbf{k})$, e.g., $~k_z^3$, are in principle allowed. Such terms preserve the $C_4$ symmetry and do not open a gap but modify the surface states. The Hamiltonian $H_1(\mathbf{k})$ has also been shown to exhibit a transition between two phases with different mirror Chern numbers\cite{Bednik_PhysRevB.98.045140}, as $B(\mathbf{k})$ passes through zero, corresponding to going from compressive to tensile strain. We are ignoring high-order terms here as the fits to the calculated DFT bandstructure did not require them, and focus on compressive strain that breaks the $C_4$ symmetry.  
%is equivalent to applying strain to the system in such a way that $C_{4v}$ symmetry about the tetragonal $c$-axis is preserved. If this symmetry is broken (eg by strain in the plane perpendicular to the $c$-axis). We consider the lowest-order term is $B_1k_z$. 
With $b_1\not=0$, a gap will open up at the Dirac nodes, and a mass will be introduced. Although more complex functional forms of $B(\mathbf{k})$ could be considered\cite{Wang_PhysRevB.88.125427}, we have only considered the simplest form that breaks $C_4$ symmetry. While parameters for the Hamiltonian in Eq.~(\ref{eqn:H4}) have been published earlier\cite{Wang_PhysRevB.88.125427,jeon_nat_mater_2014} we obtain our parameters by fitting the bulk dispersion relation along different momentum directions to the electronic band structure calculated within DFT for both unstrained and strained cases. 

%For the DFT calculations, we started with the experimental lattice parameters ($a=b=12.633$ \AA\;and $c=25.427$ \AA) and internal atomic positions  for a centrosymmetric Cd$_3$As$_2$ (Space group I4$_1$/acd).~\cite{Ali:2014} We used the pseudopotential projector-augmented wave method~\cite{GKresse:1999} implemented in the Vienna {\em ab initio} simulation package~\cite{GKresse:1996,GKresse:1993} to optimize the lattice parameters (with fixed atomic positions) for both unstrained and a strain perpendicular to the $a$-axis. The obtained ratios of $b^{\prime}/b$ and $c^{\prime}/c$) of the lattice parameters, together with a pre-set $a^{\prime}/a$, were then used to rescale the experimental lattice constants, on top of which the electronic band structure was calculated with the full-potential linearized augmented plane wave (FP-LAPW) method as implemented in the Wien2k code.~\cite{PBlaha:2020} A generalized gradient approximation~\cite{JPerdew:1996} was used. The spin-orbital coupling was included in the FP-LAPW calculations. For the presented results, a 0.7\% $a$-direction compressive strain was used.

 \begin{figure*}[hbt!]
    \begin{center}
       \includegraphics[width=15cm]{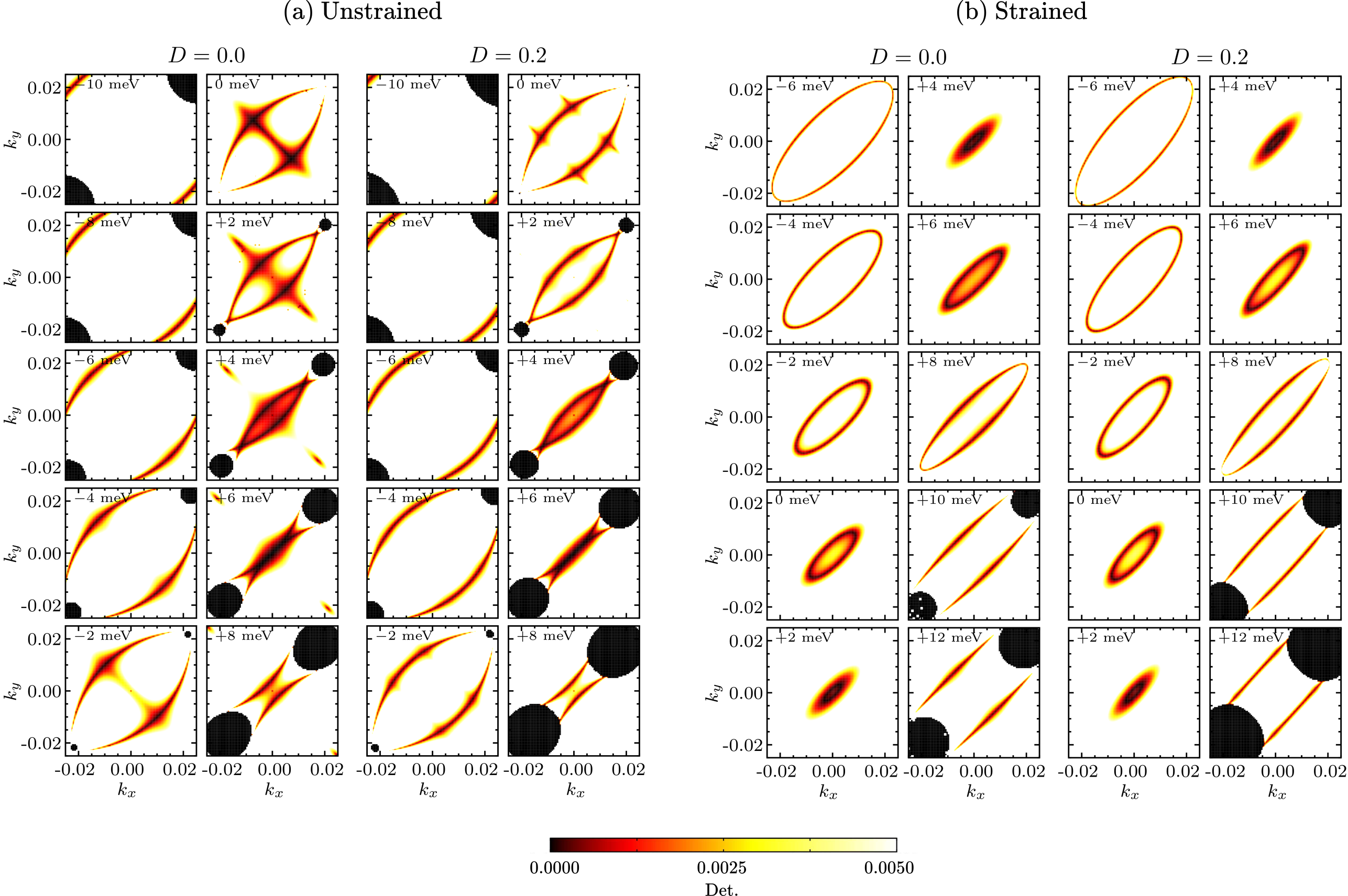}%{t1.pdf}
       \caption{Fermi arc states for a semi-infinite slab for the unstrained (a) and the strained case at -0.7$\%$ (b) and for two values of the inversion-symmetry-breaking parameter, $D=0$~ eV/{\AA} and $D$=0.2~eV/{\AA}. Units of $k_x$ and $k_y$ are \AA$^{-1}$. The color scale indicates the numerical value of the determinant from the boundary condition. Inside each panel, we indicate the energy in meV with respect to the Dirac points (in the unstrained case) or to the gap center (strained case). The black areas correspond to the bulk states. The momenta $(k_x,k_y)$ are perpendicular to the [112] crystallographic direction.}
       \label{fig:semi-infinite}
    \end{center}
 \end{figure*}

\begin{figure*}[hbt!]
    \begin{center}
       \includegraphics[width=15cm]{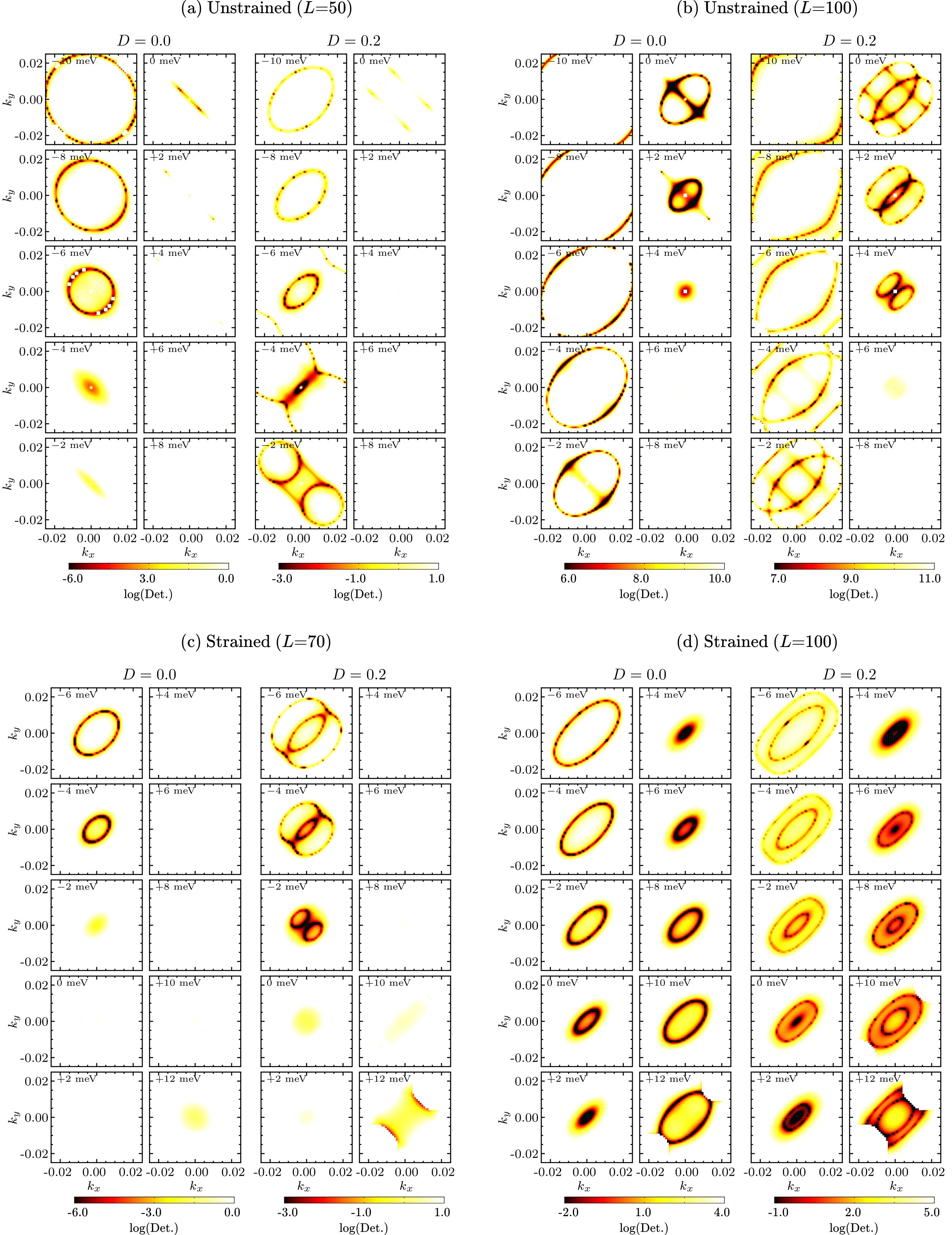}
       \caption{Fermi arc states for several thin films in unstrained Cd$_3$As$_2$ (a),(b) and for the strained case at -0.7$\%$ (c),(d). For each of these cases we analyze two thicknesses ($L=50,100,$~{\AA} for the unstrained case and $L=70,100,$~{\AA} for the strained one), and two values of the inversion-symmetry-breaking parameter $D=0.0, 0.2$~eV/{\AA}. In color scale we plot the logarithm of the determinant of the boundary condition generalized to the thin film case. Inside each panel, we indicate the energy in meV with respect to the Dirac points (in the unstrained case) or to the gap center (strained and intermediate cases). The momenta $(k_x,k_y)$ are perpendicular to the [112] crystallographic direction.}
       \label{fig:thin_film}
    \end{center}
\end{figure*}
%\noindent{\em{Fit to the DFT calculations}}:
Details about the DFT calculations, the fitting process, and the obtained parameters $A$, $C_0$, $C_1$, C$_2$, $M_0$, $M_1$, $M_2$, and $b_1$ are presented in the Supplemental Material\footnote{See Supplemental Material}. We consider two cases: (i) the relaxed structure with the experimental lattice parameters (that we call ``unstrained"), and (ii) a case with a -0.7$\%$ compressive strain applied along the $a$ crystallographic axis (``strained''). We solve for the Fermi arc states by solving the determinantal equations that result from imposing the boundary conditions\cite{hashimoto2017boundary} on a semi-infinite or finite-thickness slab (see Supplemental Material for details). 
We note that for the unstrained case, in which the $C_4$ symmetry around the crystallographic $c$ axis is preserved, the band structure is gapless and has two Dirac nodes located at momenta $k^0_z=\pm\sqrt{-M_0/M_1}$. In this case the result of the fit yields $b_1=0$, as expected. As soon as this parameter is non-zero, a non-trivial gap opens up in the bulk band structure. %Moreover, in case tensile strain is applied, a gap opens up as well but the system becomes a trivial insulator instead. 

\noindent{\em Semi-infinite slab}:
In Fig.~\ref{fig:semi-infinite} we show the momenta of the Fermi arc states in a semi-infinite slab of Cd$_3$As$_2$ oriented along the [112] crystallographic direction for different energies. %We show several cuts at fixed energies in which we plot the Fermi arc states. 
We can see the evolution of the Fermi arc states in four different cases: two different values of the parameter $D$ (which breaks inversion symmetry) for strained and unstrained systems. The value of $D$ is added to the parameters obtained from the fits to the DFT bulk bands.  %(which breaks the $C_4$ symmetry around the crystallographic $c$ axis) that have been obtained fitting the DFT band structures.

As $D$ is increased from zero, inversion symmetry is broken. This lifts the Kramers degeneracy in the topologically trivial bulk bands. Interestingly, the Fermi arc states remain rather unchanged. For the unstrained case, we find a gapless bulk energy dispersion with two Dirac cones along the $(k_x',k_y')$ direction, perpendicular to the [112] crystallographic direction. The main effect of strain is to open a gap in the bulk bands, while the Fermi arc states remain rather unchanged as a function of $D$ in the case of semi-infinite slabs. It has been shown that Fermi arc states can survive under certain symmetry breakings provided the bulk states maintain mirror symmetries~\cite{Bednik_PhysRevB.98.045140, kargarian_pnas_2016}. This is the case here, in which the two mirror symmetries with planes along the lines $k'_y=\pm k'_x$ are preserved, despite the finite values of $b_1$ and $D$, thus ensuring the presence of the surface states in the gap.~\cite{yang2014classification, Bednik_PhysRevB.98.045140, kargarian_pnas_2016}

The Fermi arc states also evolve in an interesting fashion as the bulk gap is opened. The set of two Fermi arcs that connect the two bulk Dirac cones in the unstrained gapless case only touch at the Dirac nodes forming an elliptical contour. As strain is applied and the bulk bands become gapped, this contour smoothly evolves into a closed loop that extends throughout the gap. This leads to a Dirac cone of surface states in the gap~\cite{yi2014_sci_rep,Bednik_PhysRevB.98.045140}: Under strain the DSM transforms into a TI, and the Fermi arc states of the DSM have transformed to the TSSs of a TI. 

%\textcolor{blue}{ This needs to be made sharper because it is a key insight we can get out of this model. You'd like to be able to state unequivocally that it transformed to a $Z_2$ TI and we see the TI surface states.}

\noindent{\em Thin film limit}:
In Fig.~\ref{fig:thin_film} we show the Fermi arc states for different thicknesses $L$ as a function of the symmetry-breaking parameter $D$ and of strain. As one would expect, for large values of $L$ one recovers the same result as in the semi-infinite slab. We have confirmed this for $L=500$~{\AA} (not shown). 
%Indeed, this is what happens already for $L=500$~{\AA}, for which the Fermi arc states for both the strained and unstrained case resemble those for the semi-infinite slab shown in  Fig.~\ref{fig:semi-infinite}.

As $L$ is decreased, the Fermi arc states start to evolve until they disappear in the region of energy and momenta around the Dirac points. This happens for a thickness of around $L=20~${\AA} in the unstrained case and for a thickness below $L=50~${\AA} in the strained case. We can see in Fig.~\ref{fig:thin_film} how for the unstrained cases (a) and (b), for the same values of energy, the contours become smaller for lower values of $L$. As $L$ is reduced, these contours will eventually disappear at that particular energy. In the strained case the situation is a little bit more complex. Surface states seem also to reappear at positive energies relative to the middle of the gap [$L=70$ in Fig.~Fig.~\ref{fig:thin_film} (c)]. As the thickness is increased one eventually recovers a cone of surface states, which resembles the behavior observed for the strained case in the semi-infinite slab.

We observe also how the effect of breaking inversion symmetry by setting $D\not=0$ is very much more pronounced in thin films than in semi-infinite slabs, in which the Fermi arc state were barely changed by this perturbation. In thin films, breaking the Hamiltonian inversion symmetry and lifting the Kramers degeneracy results in two sets of Fermi arc states that intersect or enclose each other. We believe this is a general feature of the Fermi arcs in the thin film limit and not limited to the specific model of Cd$_3$As$_2$ used here. %Here we can see how for different thicknesses, including a non-zero value of this inversion-symmetry breaking parameter which lifts the Kramers degeneracy and now results in two overlapping contours of Fermi arc states. %parameter creates 2 overlapping surfaces of Fermi-arc states. 

Interestingly, the thickness at which the Fermi arc states vanish completely is much smaller, around one order of magnitude, than the thinnest films fabricated so far ($\sim100$~\AA), for which the dispersion of the Fermi arc states is practically identical to that in the semi-infinite slab. This allows to confirm that the Fermi arc states survive down to the extreme thin-film limit. 

Moreover, the surface states survive in both unstrained and strained thin films, this is under broken inversion symmetry and broken $C_4$ symmetry along the crystallographic $c$-axis. It has been observed that the low-field-limit transport properties in Cd$_3$As$_2$ exhibit a strong anisotropy when the compound is grown along the [112] crystallographic direction~\cite{Goyal_PhysRevMaterials.3.064204}. The surface states will participate in transport and display the same type of symmetry as can be seen from the shape of the Fermi arc states, in particular in the range of energies immediately above and below the Dirac nodes. We speculate that this anisotropy of the surface states could be linked to the anisotropy seen in transport, although a more detailed and thorough transport study is needed.

In addition, the strong particle-hole asymmetry and anisotropy of the surface states, especially in the thin film limit, will give rise to a rich variety of low-temperature transport properties. For example, with the chemical potential fixed, only Fermi arc states below it contribute to transport as the temperature is decreased, with the consequence that transport properties would smoothly develop a robust non-trivial contribution (as seen experimentally~\cite{Goyal_APL6_Cd3As2_thin_films}). Also, by effectively moving the chemical potential (by gating or by engineering heterostructures) the presence of surface states would become noticeable. Indeed, this has been observed experimentally~\cite{Galletti_PhysRevB.97.115132}.

In general, the simplicity of the Dirac nodes and the robustness of the Fermi arc states in the thin-film limit and under different symmetry breakings investigated here, as well as  experimentally~\cite{Goyal_PhysRevMaterials.3.064204, Galletti_PhysRevB.97.115132, Galletti_PhysRevMaterials.2.124202, Schumann_PhysRevLett.120.016801}, makes this material an ideal candidate for engineered heterostructures that probe transport that leverage the properties of Fermi arc states, such as the inverse Edelstein effect~\cite{Zhang_Heinonen_PhysRevLett.123.187201}, or spin-torque effects\cite{li2019magnetization}.

We have also observed that varying the parameters of the model to values closer to those from Ref.~\onlinecite{Wang_PhysRevB.88.125427}, the results change. In particular the Fermi arc states remain in the thin film limit down to much thinner thicknesses ($\sim 5$ \AA) whereas the behavior in the semi-infinite case is qualitatively similar,  apart from an energy shift and a reduction in the Lifshitz energy: the Fermi arc states are still present despite a relatively large difference in the parameters of the model, although its dispersion and low-energy features are different. This implies that the main results are not an artifact of our fit, although specific details in the thin-film limit are affected by the parameter choice.  %and that are material dependant, which makes this a great approach to model to explore Fermi arc states in other DSM.

%\section{Summary and conclusions}
In summary, 
we have studied the behavior of Fermi arc states in Cd$_3$As$_2$ both in a semi-infinite slab and in the thin film limit under strain and broken inversion symmetry. These Fermi arc states (arising from the linear combination of evanescent states satisfying the proper boundary conditions and protected by the presence of mirror symmetries in the bulk bands~\cite{yang2014classification, Bednik_PhysRevB.98.045140,kargarian_pnas_2016}) survive to a film thickness below 50~\AA, smaller than presently reached experimentally, about 100~{\AA}~\cite{Goyal_APL6_Cd3As2_thin_films}, and also survive under compressive strain, the only effect of which is to open up a gap in the bulk bands both in the semi-infinite slab and in thin films (as has been observed experimentally~\cite{Goyal_PhysRevMaterials.3.064204}). Our findings are consistent with several experimental observations~\cite{Goyal_PhysRevMaterials.3.064204, Galletti_PhysRevB.97.115132, Galletti_PhysRevMaterials.2.124202, Schumann_PhysRevLett.120.016801}. Although we used specifically Cd$_3$As$_2$ as a model system, our approach using ab-initio determined parameters in conjunction with this four-band model can be generally applied to other candidate DSMs and WSMs. Breaking inversion symmetry in the Hamiltonian results in two sets of Fermi arc states in the thin film limit. We speculate that this is a general feature that will occur in other DSMs and is in principle observable. %This strongly suggests that the and opens the possibility for the use of our approach in other potential candidate DSMs and WSMs.
%%

%\acknowledgments
This work was supported the Center for the Advancement of Topological Semimetals, an  Energy  Frontier  Research  Center  funded  by  the  U.S.  Department  of  Energy  Office  of  Science,  Office  of  Basic  Energy Sciences.
\bibliography{Cd3As2}
\end{document}